\documentclass{aa}
\usepackage{latexsym}
\usepackage{epsfig}
\newcommand{\radm}{rad~m$^{-2}$} 
\newcommand{\pl}{\parallel}
\newcommand{\vep}{\varepsilon}

\newcommand{\srm}{\mbox{$\sigma_{\scriptscriptstyle\rm RM}$}}
\newcommand{\dg}{\mbox{$^{\mbox{\tiny o}}$}}
\begin{document}

\title{Properties of the warm magnetized ISM, as inferred
  from WSRT polarimetric imaging}

\author{M. Haverkorn\inst{1}
       \and
       P. Katgert\inst{2} 
       \and 
       A.G. de Bruyn\inst{3,4} }

\offprints{M. Haverkorn}

\institute{Leiden Observatory, P.O.Box 9513, 2300 RA Leiden, the
           Netherlands\\ 
	   (Current address: Harvard-Smithsonian Center for
           Astrophysics, 60 Garden Street MS-67, Cambridge MA 02138,
           USA)\\
	   \email{mhaverkorn@cfa.harvard.edu}
	   \and
           Leiden Observatory,
           P.O. Box 9513, 2300 RA, Leiden, the Netherlands
               \email{katgert@strw.leidenuniv.nl}
          \and
           ASTRON, 
	   P.O. Box 2, 7990 AA Dwingeloo, the Netherlands
               \email{ger@astron.nl}
	   \and
	    Kapteyn Institute,
	    P.O. Box 800, 9700 AV Groningen, the Netherlands
             }

\date{Received, accepted}

\abstract{
We describe a first attempt to derive properties of the regular and
turbulent Galactic magnetic field from multi-frequency polarimetric
observations of the diffuse Galactic synchrotron background. A
single-cell-size model of the thin Galactic disk is constructed which
includes random and regular magnetic fields and thermal and
relativistic electrons. The disk is irradiated from behind with a
uniform partially polarized background. Radiation from the background
and from the thin disk is Faraday rotated and depolarized while
propagating through the medium. The model parameters are estimated
from a comparison with 350~MHz observations in two regions at
intermediate latitudes done with the Westerbork Synthesis  Radio
Telescope. We obtain good consistency between the estimates for the
random and regular magnetic field strengths and typical scales of 
structure in the two regions. The regular magnetic field strength
found is a few $\mu$G, and the ratio of random to regular magnetic
field strength $B_{ran}/B_{reg}$ is 0.7 $\pm$ 0.5, for a typical scale
of the random component of 15 $\pm$ 10 pc. Furthermore, the regular
magnetic field is directed almost perpendicular to the line of sight.
This modeling is a potentially powerful method to estimate the
structure of the Galactic magnetic field, especially when more
polarimetric observations of the diffuse synchrotron background at
intermediate latitudes become available.
}

\titlerunning{Properties of the warm magnetized Galactic ISM}
\maketitle

\section{Introduction}
\label{s3:intro}

Since the first interpretation of small-scale structure in the
linearly polarized component of the Galactic synchrotron background as
being due to Faraday rotation (Wieringa et al.\ 1993), many
observations of the Galactic synchrotron emission have shown intricate
structure in polarization on many scales, often unaccompanied by
structure in total power. Although it has been recognized that the
Faraday rotation and depolarization of the polarized synchrotron
emission are due to small-scale fluctuations in the Galactic magnetic
field, thermal electron density and/or the line of sight, a {\it
  quantitative} description of this structure has proven to be extremely
difficult. This is because depolarization arises both along the line
of sight and in the plane of the sky (i.e.\ within the telescope beam),
whereas the rotation measure $RM$ is an integral over the line of
sight of thermal electron density $n_e$ and Galactic magnetic field
along the line of sight $B_{\pl}$: $RM [\mbox{rad~m}^{-2}] = 0.81 \int \,
n_e [\mbox{cm}^{-3}] \,B_{\pl} [\mu\mbox{G}]\,ds[\mbox{pc}]$.

Use of the diffuse Galactic synchrotron emission as a tracer of the
Galactic magnetic field is complementary to the magnetic field
estimates derived from pulsars (e.g.\ Rand  \&  Kulkarni 1989, Ohno \&
Shibata 1993) and extragalactic radio sources
(e.g.\ Simard-Normandin \& Kronberg 1980, Clegg et al.\ 1992, Brown
et al.\ 2003), in that the diffuse background can provide a continuous
field of RMs on scales from the field size down to the
resolution of the observation. Therefore this is a unique method to
infer scales and amplitudes of fluctuations in the Galactic magnetic
field and the electron density.

The RM of the synchrotron background has been used to
determine the nature of distinct objects (Gray et al.\ 1998, Uyan\i
ker \& Landecker 2002, Haverkorn et al. 2003a), to estimate the
uniform component of the Galactic magnetic field (Haverkorn et al.\
2003b) or the magnetic field strength and structure in supernova
remnants (Uyan\i ker et al.\ 2002), and to infer the polarization
horizon (Uyan\i ker et al.\ 2003). However, to the authors' knowledge,
this is the first attempt to derive the turbulent component of the
magnetic field (not associated with any discrete structure) from the
diffuse synchrotron background.

In this paper, we present a simple model of the Galactic thin disk as
a synchrotron emitting and Faraday-rotating medium, consisting of
cells with a certain electron density and magnetic field. We
compare the model predictions with observed properties of the linear
polarization in two fields observed with the Westerbork Synthesis
Radio Telescope (WSRT), to derive 
estimates for several physical parameters of the warm ISM. In
Sect.~\ref{s3:obs} we discuss the WSRT polarization observations that
will be used to compare to the model. Section~\ref{s3:depth} describes
depth depolarization in a layer that contains both
synchrotron-emitting and Faraday-rotating material.  We discuss in
Sect.~\ref{s3:comp} the ingredients of a model for a thin
disk, combined with a thick disk or halo providing a constant
polarized background.  In Sect.~\ref{s3:model} we will describe the
model in some detail, and how observational constraints can be used to
derive estimates for parameters like the magnetic field strength and
direction. In Sect.~\ref{s3:modobs} we apply the model to our
observations and discuss its results. Finally, we present a
summary and conclusions in Sect.~\ref{s3:conc}.

\section{The observations}
\label{s3:obs}

% ------------------------------------------------------------------
\begin{table}
  \begin{center}
    \begin{tabular}{l|cc}
                & Auriga           & Horologium \\
      \hline
      (l,b)     & (161\dg, 16\dg)  & (137\dg, 7\dg) \\
      size      & 7\dg$\times$9\dg & 7\dg$\times$7\dg \\
      FWHM      &5.0\arcmin$\times$6.3\arcmin&5.0\arcmin$\times$5.5\arcmin\\
      pointings & 5$\times$7       & 5$\times$5 \\
      bandwidth & 5 MHz            & 5 MHz \\
      frequencies & \multicolumn{2}{c}{341, 349, 355, 360, 375 MHz}\\
      \hline
    \end{tabular}
    \caption{WSRT polarization observations in the constellations
    Auriga and Horologium. Given are position and size of each region,
    the resolution, the  number of pointings used to mosaic the
    region, the frequency bandwidth and the central frequency in each band.}
    \label{t3:data}
  \end{center}
\end{table}
% ------------------------------------------------------------------
Two fields of observation in the constellations Horologium and Auriga,
described in detail in Haverkorn et al. (2003a, 2003b) are
used to estimate parameters  of the depolarization
model. Observations out of the Galactic plane are used to avoid
discrete objects like supernova remnants and H\,{\sc ii} regions,
which would skew the statistical information of the radiation that we
use. Relevant properties of the observations are given in Table~\ref{t3:data}. 
Undetected large-scale components in Stokes $Q$ and $U$ are not
thought to be important in these fields around 350~MHz (Haverkorn et al.\
2004).

\begin{figure*}
\begin{center}
\hbox{\psfig{figure=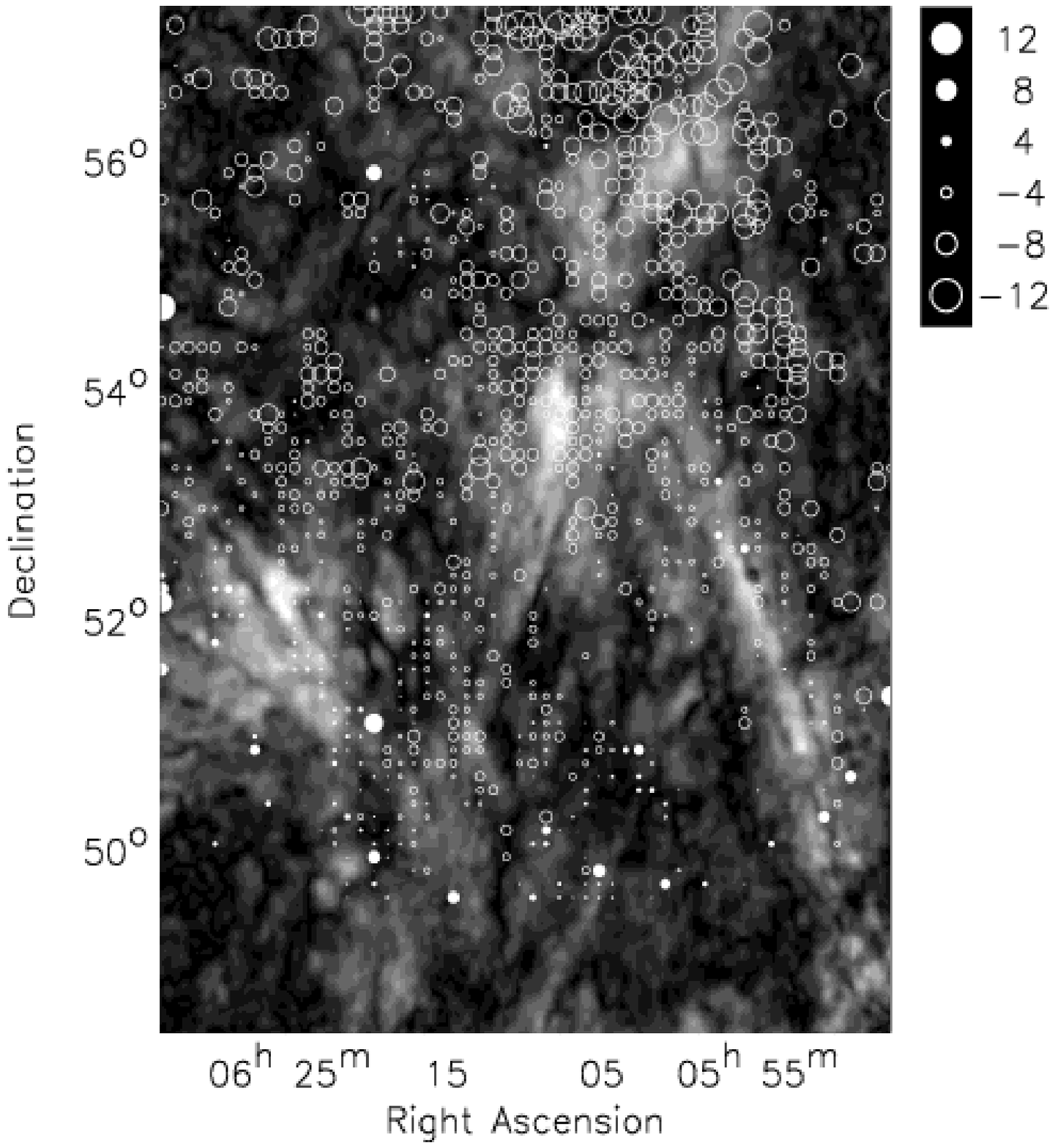,width=.49\textwidth}
      \psfig{figure=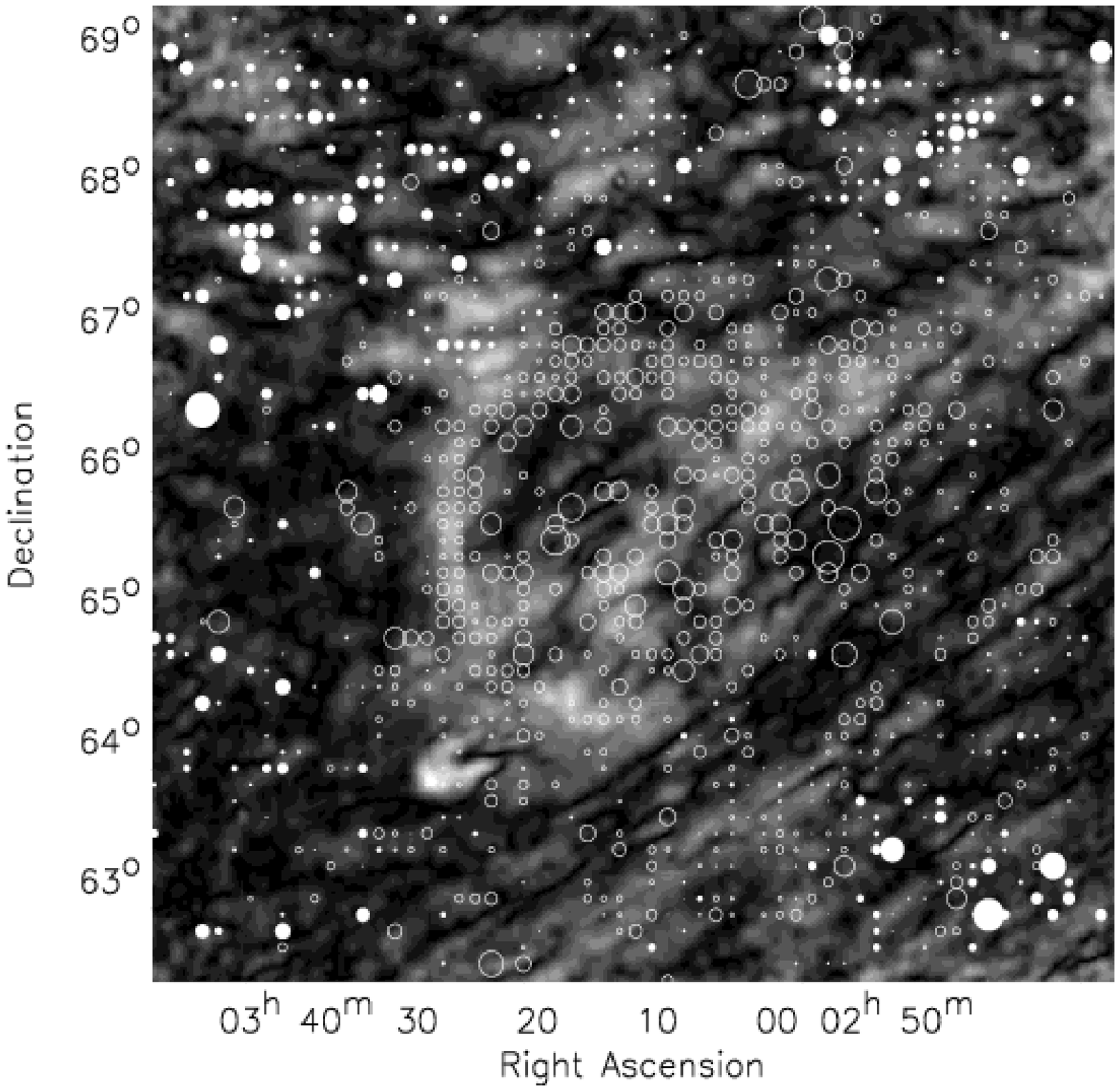,width=.53\textwidth}}
     \caption{RM maps of the observed regions in Auriga
      (left) and in Horologium (right), overlaid on polarized
      intensity at 349~MHz in grey scale. White denotes high polarized
      intensity, with a maximum of 95~mJy/beam. RMs are
      denoted by white circles, and filled circles are positive
      $RM$s. The diameter of the symbol represents the magnitude of
      $RM$, where the scaling is given in \radm. Only $RM$s for which
      $\langle P \rangle \ge 5\sigma$ and reduced $\chi^2$ of the
      linear $\phi(\lambda^2)$-relation $< 2$ are shown, and only
      every second synthesized beam.}
\label{f3:rmmap}
\end{center}
\end{figure*}

Fig.~\ref{f3:rmmap} shows the linearly polarized intensity
$P=\sqrt{Q^2+U^2}$ of the two regions in grey scale. The structure in
$P$ is uncorrelated with total intensity $I$, which does not show any
structure on scales visible to the interferometer down to noise
level. Rotation measures $RM$ were derived 
from the polarization angle $\phi = \phi_0
+ RM\lambda^2$, where the ambiguity in $\phi$ over $n \, 180\dg$ has
been taken into account (Haverkorn et al.\ 2003b). $RM$ maps are
shown as circles in Fig.~\ref{f3:rmmap}. The $RM$ in
the Auriga region (left) shows a gradient of about 1~\radm\
per degree in the direction of position angle $\theta = -20\dg$ (N
through E). 

The linear $\phi(\lambda^2)$-relation can be destroyed by
depolarization, which yields incorrect $RM$ values. Therefore, we only
consider ``reliably determined'' $RM$ values, where ``reliable'' is
defined by (a) the reduced $\chi^2$ of the linear
$\phi(\lambda^2)$-relation $\chi^2_{red} < 2$, and (b) the polarized
intensity averaged over frequency $\left<P\right> > 20$~mJy/beam
($\sim4$~--~$5\sigma$). 

%***********************************************
%***********************************************

%***********************************************
\begin{figure}
\begin{center}
\psfig{figure=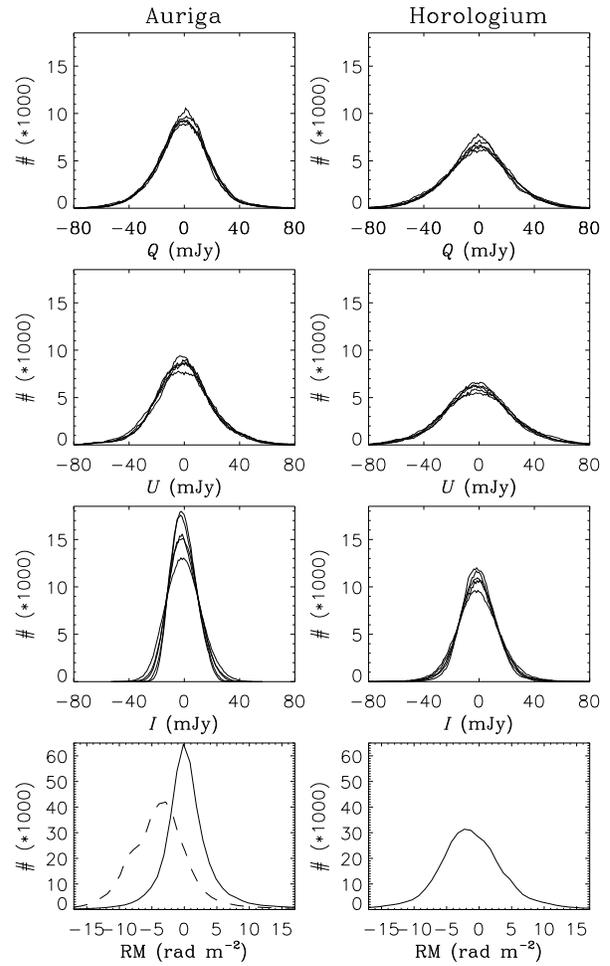,width=.45\textwidth}
      \caption{Histograms of (from top to bottom) $Q$, $U$ and $I$ for
      5 frequencies, and $RM$ for Auriga (left) and Horologium
      (right). Data of $Q$, $U$ and $I$ are 5 times oversampled, and
      only reliably determined $RM$s are included. In the solid line
      histogram of $RM$ in the Auriga region, the $RM$ gradient over
      the region is subtracted; the dashed line gives the histogram of
      the observed $RM$ including the gradient.}
\label{f3:hist}
\end{center}
\end{figure}
%***********************************************
Histograms of the distributions of Stokes parameters $Q$, $U$ and $I$,
and of $RM$ are given in Fig.~\ref{f3:hist} for Auriga (left) and
Horologium (right). In the
$I$ map, point sources were subtracted down to 5~mJy/beam. For $Q$,
$U$ and $I$, data from all five frequencies are shown in the same
plot. The $RM$ plot of the Auriga region contains the distribution of
observed $RM$ (dashed line), as well as that of $RM$ where the best-fit
linear gradient in $RM$ has been subtracted (solid line). The
statistical information along these two separate lines of sight will
be used in Sect.~\ref{s3:modobs} to infer information on the Galactic
magnetic field and correlation lengths in the warm ionized ISM. 

\section{Depth depolarization}
\label{s3:depth}

The absence of correlated structure in $I$ appears to be a general
feature: both regions discussed here show it, as do other
observations made with the WSRT at frequencies around 350~MHz
(e.g. Katgert \& de Bruyn 1999, Schnitzeler et al., in prep).  The lack
of corresponding structure in $I$ suggests that
Faraday rotation is the main process responsible for the observed
structure in polarization. 

However, rotation of the polarization angle cannot by itself cause
structure in $P$. As shown in Haverkorn et al.\ (2004), the structure
in $P$ in the observations discussed here cannot be caused by a
missing large-scale structure component. 
Instead, depolarization must be the main creator of fluctuations in
$P$. Depolarization can essentially occur in three ways: along the
line of sight, in the plane of the sky or within the frequency
band. As the latter process only causes significant
depolarization for bandwidths much wider than those used here, we will
ignore bandwidth depolarization. Depolarization along the line of
sight is referred to here as {\it depth depolarization} (a
combination of internal Faraday dispersion and differential Faraday
rotation (Sokoloff et al.\ 1998, Fletcher et al.\
2004)), as it occurs due to averaging of polarization angles along the
line of sight. In addition, a finite width of the telescope beam can
cause {\it beam depolarization} if the structure in polarization angle
is on scales 
smaller than the beam. Beam depolarization is clearly observed in the
fields discussed here, but it cannot explain the structure
on scales larger than that of the beam (Haverkorn et al. 2004).

We are therefore led to consider the situation in which
the medium that produces the Faraday rotation is mixed with a medium
that emits synchrotron radiation, which produces structure in $P$
through depth depolarization. The low level of small-scale structure
in $I$ in all of these observations implies that the total intensity
of the synchrotron emission must be very uniform. However, the
magnetic field, and therefore the synchrotron emissivity of the thin
disk, is far from uniform (e.g. Beck et al.\ 1996, Han \& Wielebinski
2002). Therefore, the smoothness of the synchrotron total intensity
cannot be due to homogeneous synchrotron emission. Instead, the number
of turbulent cells along the line of sight must be so large that the
spatial variation in synchrotron emissivity, which is a scalar
quantity, is averaged out. Linear polarization is a vector, so that
small-scale structure is more easily preserved. Furthermore, 
total intensity is integrated over a much larger path length than
polarized intensity because it isn't depolarized. Finally, in the
second quadrant, where our observations were done, the perpendicular
component of the uniform Galactic magnetic field is believed to
dominate the component parallel to the line of sight (e.g. Beck
2001). This means that $B_{\perp}$, and therefore the emitted
synchrotron radiation, has a large uniform component.

\section{Relevant components of the ISM}
\label{s3:comp}

\subsection{Cosmic rays and thermal gas}
\label{ss3:gas}

The synchrotron intensity is the integrated non-thermal emission along
the line of sight. The intensity depends on the relativistic electron
density $n_{e,rel}$ and magnetic field perpendicular to the line of sight
$B_{\perp}$.  Beuermann et al.\ (1985) have modeled the Galactic
synchrotron emissivity $\vep$ from the continuum survey by Haslam et
al.\ (1981, 1982) at 408 MHz.  They incorporate spiral structure in
the synchrotron radiation and find two components of emission, viz.\ a
galactocentric thick and thin disk with half equivalent widths of
$h_{\vep,b} \approx 1800$~pc and $h_{\vep,n} \approx 180$~pc at the
radius of the 
Sun, respectively, for a galactocentric radius of the Sun $R_\odot
= 10$~kpc. This corresponds to an exponential scale height of
  $h_{syn,thick}  \approx 1500$~pc and $h_{syn,thin} \approx 150$~pc
scaled to a galactocentric radius of the Sun $R_\odot = 8.5$~kpc
(Beck 2001).
Note that the scale heights of the cosmic-ray electrons and
of the magnetic field must be larger than that of the synchrotron
disk, e.g. by factors 2 and 4 in case of equipartition (Beck 2001). 

The major part of the Faraday rotation is caused by the warm ionized
medium, contained in the so-called Reynolds layer (Reynolds 1989,
1991). The layer has a height of about 1~kpc, a temperature $T
\approx$~8000~K, and a thermal electron density concentrated in clumps
of $n_{e,th} \approx 0.08$~cm$^{-3}$ with a filling factor of about 20\%.

In this paper we will consider two domains in the ISM. The first
domain is the thin synchrotron disk, which coincides with the stellar
disk ($\sim 200-300$~pc), the thin H\,{\sc i} disk ($\sim$~200~pc,
Dickey \& Lockman 1990), and the disk of classical H~{\sc ii }
regions ($\sim$~60~pc). The second domain is the Reynolds layer, which
coincides with the thick synchrotron disk. There is depolarization in
both domains. The Local Bubble and Local Interstellar cloud
contributions to the $RM$ are so small that they are neglected here.

\subsection{Regular and random Galactic magnetic field}

We decompose the Galactic magnetic field in a regular large-scale
component and a random component ${\bf B = B}_{reg} + {\bf B}_{ran}$.
Estimates of the ratio of random to regular magnetic field strengths
$B_{ran}/B_{reg}$ in the literature seem to depend on the method
used. Magnetic field 
determinations using $RM$s from extragalactic sources yield
$B_{ran}/B_{reg} \approx 0.5 - 1$ (Jokipii \& Lerche 1969, Clegg et
al.\ 1992). Pulsar $RM$s indicate that $B_{ran}/B_{reg} \approx
3 - 4$ (Rand \& Kulkarni 1989, Ohno \& Shibata 1993), although this
value may be an overestimate (Heiles 1996, Beck et al.\ 2003). From diffuse
polarization measurements, Spoelstra (1984) estimates $B_{ran}/B_{reg}
\approx 1-3$, in agreement with Phillipps et al.\ (1981) who find that
$B_{ran}/B_{reg} \ga 1$. Heiles (1996) estimates an average from
several studies as $B_{ran}/B_{reg} \approx 1.6$. 

Structure in $RM$ is estimated to be present on scales at least
from 0.1 to 100~pc from observations of extragalactic point
sources (Clegg et al.\ 1992, Minter \& Spangler 1996), whereas pulsar
RMs and dispersion measures DM give cell sizes of 10 to
100~pc (Ohno \& Shibata 1993). Beck et al.\ (1999) found scale sizes
of $\sim$20~pc for the galaxy NGC~6946.

Field strengths and structure in the Galactic halo, i.e.\ in the gas
above the thin synchrotron disk at $h \ga 200$~pc, can be estimated
from observations of halos of external galaxies.  In observations of
synchrotron emission in halos of edge-on galaxies, the degree of
polarization mostly increases with distance from the plane, suggesting
a decreasing irregular magnetic field component for increasing
distance to the plane of the galaxy. 
Structure in the halo varies on much larger scales than
in the thin disk, viz.\ on scales of about 100 -- 1000~pc (e.g. Dumke
et al.\ 1995).

\section{A model of a Faraday-rotating and synchrotron-emitting layer}
\label{s3:model}

In this section we describe a simple model of a thin Galactic disk
containing cosmic rays, magnetic fields and thermal electrons,
irradiated by a uniform polarized background from the thick
synchrotron disk. We calculate the total intensity, Stokes $Q$ and
$U$, and the implied $RM$, for various assumptions about the structure
of the layer. In Sect.~\ref{s3:modobs}, we will compare
these results with the observations.

\subsection{Outline of  the  model}
\label{ss3:out}

%****************************
\begin{figure}
\begin{center}
\psfig{figure=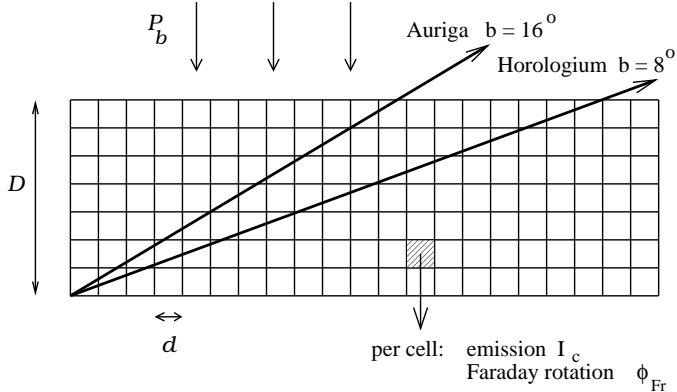,width=.5\textwidth}
\caption{Depth depolarization model on a grid of thickness $D$,
containing cells with cell size $d$. Synchrotron emission $I_c$ is
emitted in each cell, Faraday rotation only occurs in a fraction $f$
of the cells. A constant background polarization $P_b$ is also
Faraday-rotated while propagating through the medium.  Along 2 lines
of sight (Auriga and Horologium) the model is compared to
observations. }
\label{f3:mod}
\end{center}
\end{figure}
%*****************************
We describe structure in the warm gas and in the magnetic field in the
thin disk with a single-cell-size model.
Fig.~\ref{f3:mod} gives a sketch of the model and its parameters; in
addition to the cell size $d$ these are $D$, the vertical thickness of
the layer, and the synchrotron emissivity $I_c$ in each cell. The warm
ionized medium has a filling factor $f$; this is accounted for in the
model by setting the electron density to an assumed constant value
$n_e$ in a fraction $f$ of the cells along the line of sight, which
are randomly chosen. In the remaining fraction $(1-f)$ of cells, $n_e$
is set to zero as an approximation for both the hot dilute gas and the
cold neutral medium. Thus, we have made the simplifying assumption
that neither the hot nor the cold gas contribute significantly to the
$RM$.

The magnetic field in the thin turbulent disk consists of a random and
a regular component $B_{ran}$ and $B_{reg}$. The field strengths of
both components are assumed constant (but not equal).  As it is not
known how $B_{ran}$ and $B_{reg}$ in the cold, warm and hot
phases of the ISM are related, we consider two extreme cases:
\begin{itemize}
\item[A:] properties of both the random and the regular component of
  the magnetic field are identical in all phases of the ISM.
\item[B:] the random component of the magnetic field only exists in
  the turbulent warm ISM. In the cold and hot ISM, the regular
  magnetic field component predominates. The total magnetic field
  energy density is equal in all phases.
\end{itemize}
{Hence, the properties of the cells are identical in both models --
they contain $n_e$, $B_{ran}$ and $B_{reg}$. The difference between
models~A and~B is the hot/cold medium (outside the cells) which
contains $B_{ran}$  and  $B_{reg}$ (model~A), but only $B_{reg}$ in
model~B.}

In each cell, an amount of synchrotron radiation $I_c \propto
B_{\perp}^2$ is emitted, which is assumed to be 70\% linearly
polarized. In the cell, the polarized component of this emission and
all emission from behind is Faraday rotated by an amount $\phi_{Fr}$. 
 
The thick synchrotron disk serves as a background to the detailed model
of the thin disk described here. We assume that the structure in the
thick disk is on such large scales that we can approximate the
background as a uniform synchrotron emitter, producing a constant
polarized intensity $P_b$ as input to the thin disk, with uniform
polarization angle. $P_b = 0.7\;\eta_b\;I_b$, where $I_b$ is the total
intensity of the background, and $\eta_b$ is a constant
depolarization factor describing the depolarization due to the thick
disk, with $0 \le \eta_b \le 1$. Details of the emission and
propagation of the radiation can be found in Appendix~\ref{a:model}.

Each line of sight through the model grid (to be identified with
the direction of one of our fields, and corresponding to a particular
Galactic latitude) is simulated many times, by independently filling
the cells that contain the warm ISM, and by redrawing the angle that
the random component of the magnetic field makes with the line of
sight. An ensemble of such realizations, for  which we
derive the distributions of $I$, $Q$, $U$ and $RM$, simulates the many
lines of sight for which we obtain the same information in one of the
observed fields. So, only the statistical information of
these distributions is included in the model, not the discrete
structure. Each of the two observed regions separately already
provides useful constraints for the model parameters (see below), but
the combined set of constraints in Table~\ref{t3:obs} is quite
powerful by virtue of the different path lengths through the medium.

As we model many lines of sight by redrawing the same line of sight
many times for different models, beam depolarization is not included
in the models.

\subsection{The various types of model parameters}
\label{ss3:parm}

Four types of parameters, listed in Table~\ref{t3:mpar}, are used in
the model. We discuss these separately.

\paragraph{Input parameters with fixed values}

These are physical parameters which can be estimated or for which 
reasonably good estimates exist in the literature. From dispersion measures
($DM$) of pulsars in globular clusters at high Galactic latitude and
H$\alpha$ emission measures ($EM$), Reynolds (1991) derives $n_e
\approx$ 0.08~cm$^{-3}$, with a filling factor $f =$ 40\%
if the warm ionized ISM layer has a constant electron density, and
20\% if the electron density distribution is exponential. The
Beuermann et al.\ (1985) model for Galactic synchrotron radiation
predicts a half equivalent width of the thin disk of 180~pc.  We run
the model with fixed values $f=$~20\%, $D=180$~pc and $n_e=$~0.08~cm$^{-3}$
and discuss afterwards how the results would change if these parameters
were different.

The intrinsic polarization angle of the background only defines the
average angle in the final map of polarization angle. It changes the
$Q$ and $U$ maps locally, but has no influence on the distributions of
$Q$ and $U$. Therefore the value of $\phi_0$ is arbitrary and chosen
to be 0\dg.  The angles $\alpha$ and $\phi_r$ define the orientation
of the random component of the magnetic field, with respect to the
line of sight and some fixed direction in the plane of the sky,
respectively. Both are randomly drawn from uniform distributions, for
each cell.

The total intensity $I_0$ is taken from the 408~MHz all-sky survey by
Haslam et al.\ (1982). The 2.7~K contribution from the CMBR is
subtracted before these data are converted  to a frequency of 350 MHz
using a spectral index of --2.7. Approximately 25\% of the total
background temperature is due to point sources (from source counts,
Bridle et al.\ 1972), so only the remaining 75\% is included in the
model. This yields values of 34~K and 47~K for $I_0$ in Auriga and
Horologium, respectively.

The proportionality constant $C=1$ is estimated from the local
cosmic ray spectrum, see Appendix~\ref{a:c}. If strict equipartition
between cosmic rays and 
magnetic field applies, then $C$ is not constant but varies with
$B_{\perp}^2$, so that the synchrotron emission $I \propto
B_{\perp}^4$. Although equipartition is believed to hold on global
Galactic scales, it is highly uncertain if equipartition is valid at
parsec scales as well, because fluctuations in the supply rate of
cosmic rays may destroy equipartition on small scales. Therefore, the
exponent $\alpha$ of the relation $I \propto B_{\perp}^{\alpha}$ could
be between 2~and~4. Although Burn (1966) concludes that equipartition
does not influence the depolarization much, Sokoloff et al. (1998)
find that in the case of equipartition the depolarization effects can
differ by maximally 25\%. We assume $\alpha = 2$ in the model, and
discuss afterwards the change in parameter values if $\alpha > 2$.
 
\paragraph{Free input parameters} 

No external constraints are imposed on the cell size $d$. Estimates of
cell sizes from the literature range from 10~pc to about 100~pc, and mostly
probe scales that exceed the size of our fields. We probe cell sizes
from a parsec to several tens of parsecs, and find the cell
size determined in a reasonably narrow range because of the
observational constraints. 

\paragraph{Constraints determined from the observations}

% ------------------------------------------------------------------
\begin{table}
  \begin{center}
    \begin{tabular}{ccc}
      \hline \bf Parameter 
      &\bf Auriga region & \bf Horologium region \\ 
      \hline 
      $RM_0$         & --3.4 \radm & --1.4 \radm \\ 
      \srm           & 1.8   \radm & 4.3   \radm \\ 
      $\sigma_I$     & $\le$ 1.7 K & $\le$ 2.5 K \\ 
      $\sigma_{Q,U}$ & 3 K         & 4.3 K \\
      \hline
      \multicolumn{3}{l}{\bf Additional constraints:} \\
      \multicolumn{3}{l}{Distributions of $RM$, $I$, $Q$, and $U$ are
                         Gaussian}\\
      \hline
      \multicolumn{3}{c}{\mbox{}} \\
    \end{tabular}
    \caption{Values of observationally determined parameters from
    polarization maps of the Auriga and Horologium regions and other
    observational constraints for the models.}
   \label{t3:obs}
  \end{center}
\end{table}
% ------------------------------------------------------------------

As discussed in Sect.~\ref{s3:obs}, our observations yield
distributions of $I$, $Q$, $U$ and $RM$. We summarize these results in
Table~\ref{t3:obs}, in the form of the mean value of observed $RM$,
$RM_0$, and the dispersions $\sigma_{RM}$, $\sigma_I$, $\sigma_Q$ and
$\sigma_U$. In the Auriga field, the dispersion in $RM$ was derived
after subtraction of the best-fit gradient in $RM$ (see
Sect.~\ref{s3:obs}). Like in the observations, only reliably
determined $RM$s are used.

\paragraph{Model parameters that can be adjusted and optimized}

For any chosen value of cell size $d$, the model contains five
parameters that are not derived from external 
data or from the observations. These are: the parallel and
perpendicular components of the large-scale magnetic field,
$B_{reg,\pl}$ and $B_{reg,\perp}$ respectively, the strength of the
random component of the magnetic field $B_{ran}$, the 
intensity of the polarized emission from the thick disk $P_b$, and the
thick disk depolarization factor $\eta_b$ which connects $P_b$ and $I_b$.
The five parameters have specific dependences on the observables, as
is depicted in Table~\ref{t3:depen}. 
% ------------------------------------------------------------------
\begin{table}
  \begin{center}
    \begin{tabular}{ll}
      \hline \bf Parameter & \bf depends on \\
      \hline 
      $RM_0$         & $B_{reg,\pl}$ \\
      \srm           & $B_{reg,\pl}$, $B_{ran}$ \\
      $\sigma_I$     & $B_{ran}$, $B_{reg,\perp}$ \\
      $\sigma_{Q,U}$ & $B_{ran}$, $B_{reg,\perp}$, $P_b$ \\
      $I_0$          & $B_{ran}$, $B_{reg,\perp}$, $P_b$, $\eta_b$ \\
      \hline
    \end{tabular}
    \caption{Observables and their dependencies}
   \label{t3:depen}
  \end{center}
\end{table}
% ------------------------------------------------------------------
This makes it possible to determine definite values for most free
parameters, e.g.\ $B_{reg,\pl}$ can be determined because $RM_0$ only
depends on $B_{reg,\pl}$. The free parameters are determined as
follows (where the subscript $obs$ means the observed value): 
\begin{enumerate}
\item Set $B_{reg,\pl}$ to obtain $RM_0 = RM_{0,obs}$
\item Set $B_{ran}$ to obtain \srm~=~\srm$_{,obs}$
\item Set $B_{reg_\perp}$ to match $\sigma_I = \sigma_{I,obs}$. Then:
      \begin{itemize}
	\item If $\sigma_{Q,U} < \sigma_{Q,U,obs}$: set $P_b>0$ to
	  obtain $\sigma_{Q,U} = \sigma_{Q,U,obs}$ 
	\item If $\sigma_{Q,U} > \sigma_{Q,U,obs}$: decrease
          $B_{reg_\perp}$ and find a range of ($B_{reg_\perp}$, $P_b$)
          for which $\sigma_I < \sigma_{I,obs}$ and $\sigma_{Q,U} =
          \sigma_{Q,U,obs}$, with the additional constraint that the
          $Q$ and $U$ distributions remain Gaussian.
      \end{itemize}
\item Set $\eta_b$ to obtain the correct value of $I_0$
\end{enumerate}

%***********************************************
\begin{figure*}
\begin{center}
\psfig{figure=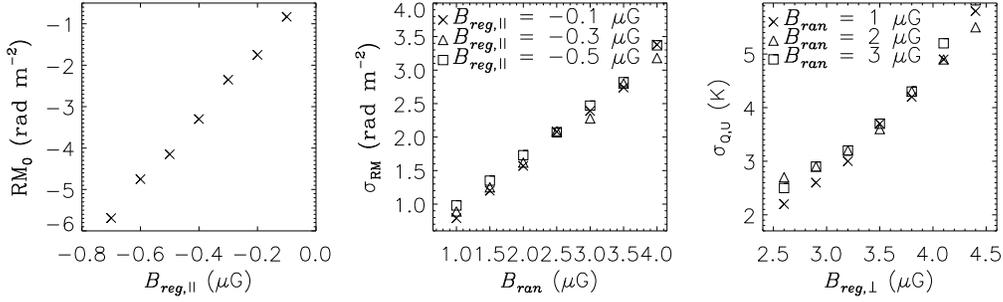,width=.75\textwidth}
      \caption{Dependences of depth depolarization parameters on
      observables in the Auriga region, for model~A. Left plot: $RM_0$
      increases with $B_{reg,\pl}$. Center plot: for fixed
      $B_{reg,\pl}$, \srm\ increases with $B_{ran}$, with only a weak
      dependence on the value of $B_{reg,\pl}$. Right plot:
      $\sigma_{Q,U}$ depends on $B_{ran}$ and $B_{reg,\perp}$.}
\label{f3:dpar}
\end{center}
\end{figure*}
%***********************************************

An example of the correlations used is given in Fig.~\ref{f3:dpar}. 
The leftmost plot shows the dependence of $RM_0$ on $B_{reg,\pl}$. As
expected, a large regular magnetic field component causes a large
non-zero mean RM (the negative value of the magnetic
field reflects the observed negative $RM_0$). The center plot gives
the change of \srm\ with $B_{ran}$ for three fixed values of
$B_{reg,\pl}$, which increases roughly linearly with $B_{ran}$ and
shows hardly any dependence of \srm\ on $B_{reg,\pl}$. Having set
$B_{ran}$ to obtain the observed value of \srm\, the right plot shows
how the observable $\sigma_{Q,U}$ depends on $B_{reg,\perp}$. The
width of the $Q$ and $U$ distribution depends slightly on the chosen
values of $B_{ran}$.

\section{Results from the model}
\label{s3:modobs}

For models A and~B as defined in Sect.~\ref{ss3:out}, the
propagation of polarized radiation through the medium is computed for
a range of values of the cell size $d$, which results in values for
the five adjustable parameters for each $d$.

The allowed cell size is well-constrained by the observations: if the
cell size is large, the number of cells is small for a given path
length and filling factor. As the number of cells with
Faraday-rotating, thermal medium can differ per line of sight, the
$RM$ distribution will not be Gaussian anymore if the cells are chosen
too large. On the other hand,
if the cell size is small, the $RM$ per cell decreases. But to obtain
a large enough \srm, the $RM$ per cell has to be rather high, so the
parameter $B_{ran}$ has to be increased to produce the observed value
of \srm. However, an increase of $B_{ran}$ increases $\sigma_I$, which
then puts an upper limit on $B_{reg,\perp}$.  To produce the observed
dispersion in $Q$ and $U$, we then need a large value for the
background polarized intensity $P_b$. If the cell size is taken too
small, $P_b$ becomes so large compared to the polarized emission in
the cells, that the distributions of $Q$ and $U$ become
distinctly non-Gaussian. Allowed values of $d$ range from approximately 1 to
60~pc, with an optimum value of about 15~pc, in good agreement with
estimates by Ohno \& Shibata (1993).  However, a cell size of 15~pc
located at the far end of the thin disk in the direction of the Auriga
region subtends an angle of more than a degree on the sky. As we
observe structure on degree to arcminute scales, smaller cells must be
present as well. Most likely, a power law spectrum of turbulence is
present on these scales (Clegg et al.\ 1992, Armstrong et al.\ 1995,
Minter \& Spangler 1996). An extension of the model including a power
law spectrum of cell sizes would increase the depolarization along the
line of sight, thereby decreasing the random magnetic field strength
needed.

%***********************************************
\begin{figure}
\begin{center}
\psfig{figure=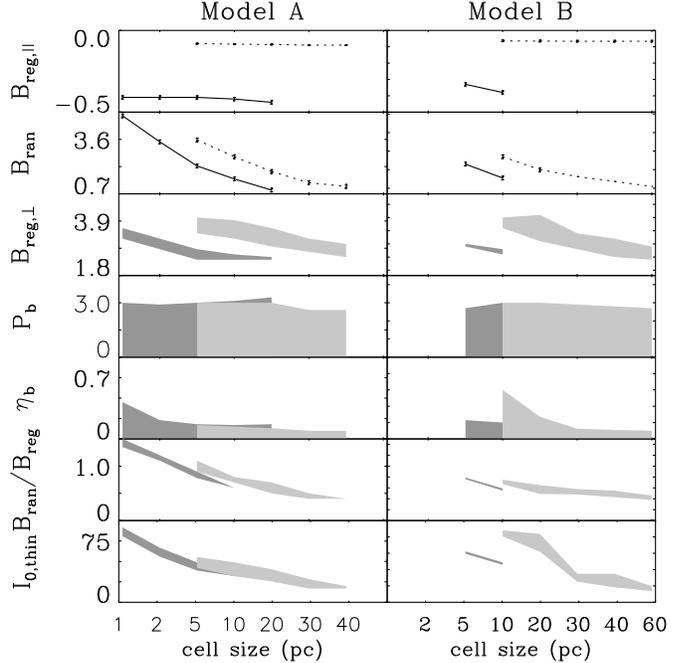,width=.5\textwidth}
      \caption{Allowed ranges of parameters for model~A (left) and~B
               (right) for different values of cell sizes. The
               magnetic field values are given in $\mu$G, 
               the polarized brightness temperature of the background
               $P_b$ in K and $I_{0,thin}$ is the percentage of the
               total emissivity that originates in the thin disk.  The
               lines in the upper plots show values found for
               $B_{reg,\pl}$ and $B_{ran}$ for the Auriga region
               (solid line) and the Horologium region (dotted
               line). All plots below those show allowed ranges in
               parameters for the Auriga region (dark) and Horologium
               (light).}
\label{f3:resmod}
\end{center}
\end{figure}
%***********************************************

We now summarize the result of the comparison between the models and
the observations. The cell sizes probed in the modeling were 1, 2, 5,
10, 20, 30, 40 and 60~pc, although for model~B, only cell sizes above
5~pc are allowed, and smaller cell sizes are allowed for the Auriga
region than for Horologium.

Fig.~\ref{f3:resmod} shows the allowed ranges of parameters in the two
regions for models A (left) and~B (right). The upper plots show values
of the obtained parallel regular field $B_{reg,\pl}$, where the solid
line denotes Auriga and the dotted line Horologium. $B_{reg,\pl}$ is
 $-0.42\pm0.02~\mu$G for Auriga and $-0.085\pm0.005~\mu$G for
Horologium in model~A, and $-0.35\pm0.01$~$\mu$G and
$-0.065\pm0.005~\mu$G respectively in model~B, where the errors are
estimated from the spread in observed values.  $B_{reg,\pl}$ hardly
depends on cell size. The next plots down show $B_{ran}$, where
the best value is about 1~$\mu$G for large ($\ga 5$~pc) cell sizes for
both the Auriga and the Horologium region, and increases to
$\sim~4~\mu$G for cell sizes of a parsec.

For the remaining parameters $B_{reg,\perp}$, $P_b$, and $\eta_b$ only
parameter ranges could be determined instead of definite values, given
in Fig.~\ref{f3:resmod} by a dark grey region for Auriga and light grey for
Horologium. The edges of the shaded areas are not sharp as drawn but
have a considerable uncertainty. The allowed parameter range should be
read more as an indication of possible parameter values than as ranges
with sharp boundaries. Moreover, the discrete edges at a certain cell
size only indicate that the next probed cell size could not produce
parameters in agreement with the observations.

The perpendicular component of the regular magnetic field
$B_{reg,\perp}$ is approximately $2.8\pm0.5~\mu$G in Auriga and
$3.2\pm0.5~\mu$G in Horologium for both models. The intensity of the
polarized background $P_b$ varies between 0.1 and 3~K, with a best
estimate of about 1.5~$\pm$~1.0~K. The depolarization factor $\eta_b =
P_b/(0.7 I_b)$ of the thick disk ranges from almost zero to 0.6
with a best estimate of about 0.15~$\pm$~0.1. 

Below that, $B_{ran}/B_{reg}$ is given, which
varies between 0.5 and 1.5 but is mostly smaller than one. The bottom
plots show $I_{0,thin}$, the percentage of the total synchrotron
emission generated in the thin disk, to be between 20\% and 75\%, and
decreasing for larger cell sizes.

Having determined these parameter ranges, we vary the filling factor
$f$, thermal electron density $n_e$ or thickness $D$ while keeping all
earlier determined parameters fixed, to gauge their influence on the
model output parameters.  A filling factor $f 
\la$~5~--~10\% is not allowed in either model: large cell sizes give a
non-Gaussian $RM$ distribution, and small cell sizes yield too high a
background polarization to keep $Q$ and $U$ Gaussian. No upper limit
can be given for the filling factor, and $B_{ran}/B_{reg}$
decreases with a factor two for $f = 1$.  For varying thermal electron
density, a low $n_e$ $\la 0.03$~cm$^{-3}$ dictates such a high
$B_{ran}$ that the ratio $B_{reg,\perp}/P_b$ becomes so low that $Q$
and $U$ become distinctly non-Gaussian. High electron densities are
allowed in the models but the random magnetic field drops to very low
values ($B_{ran} \la 0.15~\mu$G for $n_e \ga 0.1$~cm$^{-3}$).  A lower
limit to the thickness of the thin disk is about 100~pc, again no upper
limit can be set. The difference between a thickness of 150~pc or
180~pc is negligible.

We checked the influence of the assumption of equipartition between
magnetic field and cosmic rays. If $I \propto
B_{\perp}^4$ instead of $I \propto B_{\perp}^2$, the upper limit to
structure in $I$ becomes much more stringent. Therefore, model~A will
no longer produce any solutions that agree with the observables. In
model~B solutions are found with cell sizes 10 and
20~pc, and $B_{reg,\perp}$ much lower, about 0.5~$\mu$G. Other
parameters are comparable to the case where $I \propto B_{\perp}^2$.

Finally, it should be mentioned that the mean values of the
distributions of $Q$ and $U$, which are large-scale components that
are not observable with an interferometer, are lower than 1.2~K in all
models for all parameters. These large-scale components are negligibly
small, in agreement with Haverkorn et al.\ (2004).

\subsection{Discussion of the resulting model parameters}

A basic first conclusion is that the values obtained in the two
regions roughly agree, even though the Auriga and Horologium regions
have different input parameters and a different line of sight through
the medium.  The regular magnetic field components approximately agree
in the two regions ($B_{reg} \approx 3~\mu$G). 

From the deduced depolarization factor $\eta_b = 0.15$, we can estimate the
halo magnetic field, assuming a constant background. Around 350~MHz,
depth depolarization of a uniform medium to 0.15 times the original
polarization is caused by $RM$ $\approx 3 - 5$~\radm\ (Burn 1966),
which indicates a value of $B_{\pl} \approx 0.1\mu$G for a height of
the Reynolds layer of 1~kpc and $n_e \approx 0.05$~cm$^{-3}$ in the
halo. So the disk magnetic field could persist with only slight
attenuation throughout the Reynolds layer, as was suggested earlier by
Han et al.\ (1999).

Our estimate of $B_{ran}/B_{reg} < 1$ is somewhat lower than most of the
estimates from the literature discussed in Section~\ref{s3:comp}. This
may be due to several factors.  First, random magnetic field structure
on scales larger than our field of view ($\sim 7\dg\times 9\dg$) will
be interpreted as regular field in our analysis. Secondly, it could be
the result of selection, as our observational fields were chosen for
their high polarized intensity, which in our model automatically
implies a modest random magnetic field. Finally, the observations are
in the second Galactic quadrant, so we probe mostly the inter-arm
region between the Local and the Perseus arms, where $B_{ran}/B_{reg}$
is smaller than in the average ISM including spiral arms (e.g. Indrani
and Deshpande 1998, Beck 2001).

The emission in the thin disk $I_{0,thin}$ is also estimated by Beuermann et
al.\ (1985) in their standard decomposition of $I_0$ into thin and
thick disk contributions. According to their model, only about 20 to
at most 35\% of $I_0$ is generated in the thin disk and the nearest
180~pc of the thick disk. Furthermore, Caswell (1976) estimated the
synchrotron emissivity from a survey with the Penticton 10~MHz array
as 240~K~pc$^{-1}$ at 10~MHz. Rescaled to 350~MHz, this gives a total
emission from the thin disk of 10.6~K in the Auriga region and 21~K in
the Horologium region. Roger et al.\ (1999) estimate from the 22~MHz
survey performed with the DRAO 22~MHz radio telescope an emissivity of
about 55~K~pc$^{-1}$ for two HII regions in the outer Galaxy, out of
the Galactic plane. Their results give estimates of the emissivity in
the thin disk which are approximately twice as high as the estimates
from the Caswell survey. Due to the large uncertainty in the
emissivity, $I_{0,thin}$ does not put a strong constraint on the model
parameters.

\subsection{A ``polarization horizon''?}

%***********************************************
\begin{figure}[t]
\centering
\psfig{figure=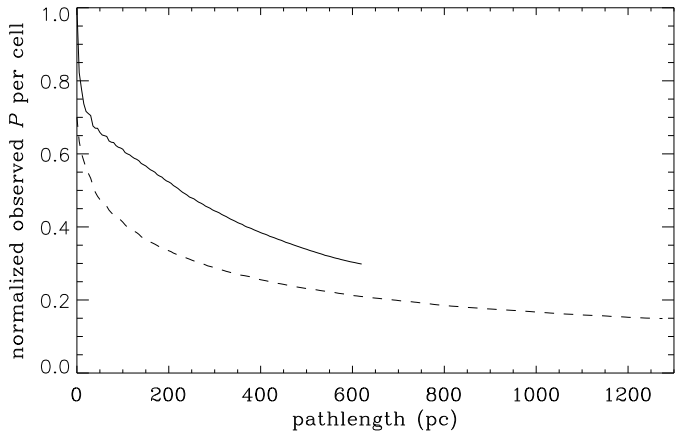,width=.5\textwidth}
\caption{Observed fraction of polarization originating closer than a certain
      distance against the path length, for the Auriga region (solid
      line) and the Horologium region (dashed line).}
\label{f3:polhor}
\end{figure}
%***********************************************

A ``polarization horizon'' is defined as a distance beyond which
(most of the) emitted polarized emission is depolarized when it
reaches the observer 
(Landecker et al.\ 2001). This can be due to beam depolarization, when
the angular scale of the structure in the polarized emission becomes
smaller than the synthesized beam at a certain distance. If the
smallest scales in the observed regions are about a parsec, the angle
of these scales on the sky becomes smaller than the beam at a distance
of about 700~pc. Spoelstra (1984) derived the polarization horizon
from comparison of radio continuum data at five frequencies from
408~MHz to 1411~MHz (Brouw \& Spoelstra 1976) with starlight
polarization. He found a distance to the origin of the polarized radio
emission of 625~$\pm$~125~pc in the direction of our fields.

Furthermore, the resulting degree of polarization decreases for
increasing pathlength through a rotating and emitting medium. However,
as depolarization is a process occurring in a telescope and not in the
medium itself, it cannot be determined what distance the remaining
polarized radiation originated. We can only estimate the decrease in
polarization as a function of path length. In our model, we build
up a line of sight by adding cells one by one, starting at the
observer. The radiation from each added cell is Faraday-rotated by all
warm ionized material in front of it. The observed degree of
polarization after addition of each cell is given as a function of the
line of sight built up until that particular cell in
Fig.~\ref{f3:polhor}, for the Auriga (solid line) and Horologium
(dashed line) regions, for model~A with a cell size of 5~pc. Even for
the total path length through the thin disk, still a fairly large
fraction (20\%) of the polarized emission can be observed. Therefore,
depth depolarization alone cannot produce a true horizon, but
attenuates the polarization gradually with path length. 

From both arguments, we estimate a distance of about 600 to
700~pc as the critical path length (``polarization
horizon''). Polarized radiation traveling along significantly larger
path lengths than this is expected to be largely depolarized.

\section{Summary and conclusions}
\label{s3:conc}

Depth depolarization, the depolarization process along the line of
sight in a medium of synchrotron-emitting and Faraday-rotating
material, is the dominant cause of structure in polarized
intensity which is unrelated to total intensity fluctuations. 

We modeled the effect of depth depolarization with a simple model of
the Galactic ISM 
consisting of a layer of cells containing random and regular magnetic
field  $B_{ran}$ and $B_{reg}$, and thermal electron density $n_e$ in
a fraction $f$ of the cells (mimicking the filling factor $f$). This
layer corresponds to the Galactic thin disk 
with small-scale structure in the magnetic field. The Galactic thick
disk or halo is modeled by a constant background $P_b$, with a certain
constant depolarization denoted by the factor $\eta_b$. We vary cell
size, magnetic field and background to obtain a range of models that
comply to the observational constraints, i.e. yield the correct width,
mean and shape of the distributions of $Q$, $U$ , $I$ and $RM$.

The results can be summarized as follows: the allowed cell size is
constrained to be in the range of 1 to 60~pc, with a best estimate of
15~pc. The regular magnetic field component along the line of sight
($\sim -0.4~\mu$G for Auriga, and $\sim -0.08~\mu$G for Horologium) is
much smaller than the regular magnetic field component perpendicular
to the line of sight ($\sim 2.8~\mu$G for Auriga, and $\sim 3.2~\mu$G
for Horologium), indicating that the regular magnetic field is
directed almost perpendicular to the line of sight in these
directions.  The random magnetic field component is about 1 to
3~$\mu$G in the two regions. In most of our models, the regular
component of the magnetic field was found to be higher than the random
component, with an average ratio of $B_{ran}/B_{reg} = 0.7\pm0.5$
which increases for smaller cell sizes. Estimates from the literature
tend to give larger 
ratios (0.5~--~4). This could be explained by the size of our fields
of view ($\sim$~7\dg\ in size), so that random components of the
magnetic field on scales large than the field are misinterpreted as regular
components. Furthermore, the fields of observation were selected for
their high polarization, indicating a higher regular magnetic field
component than average, and are situated in an inter-arm region, where
uniform magnetic fields tend to be higher than average.  The constant
polarized background intensity from the thick disk is about
$1.5\pm1.0$~K. 

This model forms a promising first attempt to derive properties of the
Galactic magnetic field from observed polarization and rotation
measures. In future work, the model can be expanded e.g.\ using
a power law distribution of the structure. Furthermore, cell
size, filling factor and electron density appear to be correlated
(Berkhuijsen 1999), which should be incorporated in a future version.
New observations in different directions can narrow down the
parameter space considerably.

\begin{acknowledgements}
We wish to thank R. Beck and E. Berkhuijsen for enlightening
discussions which improved the paper considerably, and Dan Harris for
helpful comments. The  Westerbork Synthesis Radio Telescope is
operated by the Netherlands Foundation for Research in Astronomy
(ASTRON) with financial support from the Netherlands Organization for
Scientific Research (NWO). MH acknowledges support from NWO grant
614-21-006.
\end{acknowledgements}

\appendix

\section{Outline of the depth depolarization model}
\label{a:model}
% ------------------------------------------------------------------
\begin{table*}
  \begin{center}
    \begin{tabular}{|ll|l|}
      \hline
      \multicolumn{2}{|l|}{\bf Input parameters with fixed values} 
      & \bf value: \\
      \hline
      $n_e$    & thermal electron density in cells & 
                 0.08~cm$^{-3}$ (Reynolds 1991) \\
      $f$      & filling factor of the warm ISM & 20\% (Reynolds 1991)\\
      $D$      & thickness of the layer with cells & 
		 180~pc (Beuermann et al.\ 1985)\\
      $\phi_0$ & intrinsic polarization angle of the background 
               & arbitrary: 0\dg\ chosen \\
      $\phi_r$ & position angle of random magnetic field
               & random per cell\\
      $\alpha$ & angle between random magnetic field and line of sight
               & random per cell \\
      $I_0$    & total intensity  & Auriga: 34~K \\
               &           & Horologium: 47~K (from Haslam et al.\ 1982)\\
      $C$      & proportionality constant between $I_c$ and $B_{\perp}^2$
	       & $C=1$, see Appendix~\ref{a:c}\\ 
      \hline  
      \multicolumn{3}{|l|}{\bf Free input parameters}\\
      \hline
      $d$      & \multicolumn{2}{l|}{cell size} \\
      \hline 
      \multicolumn{3}{|l|}{\bf Constraints determined from the observations}\\
      \hline
      $RM_0$      & \multicolumn{2}{l|}{mean rotation measure}\\  
      \srm           & \multicolumn{2}{l|}{width of $RM$ distribution}\\ 
      $\sigma_I$     & \multicolumn{2}{l|}{width of $I$ distribution}\\
      $\sigma_{Q,U}$ & \multicolumn{2}{l|}{width of $Q$, $U$ distribution}\\ 
      \hline
      \multicolumn{2}{|l|}{\bf Model parameters that can be adjusted
      and optimized} 
      & \bf set by dependence of: \\
      \hline
      $B_{reg,\pl}$   & parallel component of regular magnetic field
                      & $RM_0$ ($B_{reg,\pl}$) \\
      $B_{ran}$       & (constant) strength of random  magnetic field 
                      & \srm\ ($B_{reg,\pl}$,$B_{ran}$) \\
      $B_{reg,\perp}$ & perpendicular component of regular magnetic field
                      & $\sigma_I$ ($B_{ran}$,$B_{reg,\perp}$) \\       
      $P_b$           & polarized intensity of background 
                      & $\sigma_{Q,U}$ ($B_{ran}$,$B_{reg,\perp}$,$P_b$)\\
      $\eta_b$        & factor for depolarization of background
                      & $I_0$ ($B_{ran}$,$B_{reg,\perp}$,$P_b$,$\eta_b$) \\
      \hline \hline
      \multicolumn{3}{|l|}{\bf Additional constraints:} \\
      \hline
      \multicolumn{3}{|l|}{background depolarization factor $0\le\eta_b\le1$}\\
      \multicolumn{3}{|l|}{Number of cells $N = L / d$, while $N f$ cells 
	                   determine the shape of $RM$ distribution}\\
      \multicolumn{3}{|l|}{$B_{reg,\perp}$/$P_b$ determines shape of 
	                   $Q$, $U$ distribution}\\ 
      \hline
      \multicolumn{3}{c}{\mbox{}} \\      
    \end{tabular}
    \caption{The first set of parameters is determined from the
    literature or can be arbitrarily chosen. The second set is varied
    in the models, and the third set of parameters is set by our
    observations. The last set are those parameters of the ISM that
    can be estimated from the models, followed by the input parameters
    from the categories above. In parentheses the model parameters
    that they depend on.}
    \label{t3:mpar}
  \end{center}
\end{table*}
% ------------------------------------------------------------------

The synchrotron radiation emitted in each cell is $I_c \propto
B_{\perp}^2$. This emission, and the emission from each cell further
away along the line of sight and from the background passing through
the cell, is Faraday rotated by an amount $\phi_{Fr}$.  So in each
cell:  
\begin{eqnarray}
  I_c       &=& \frac{C}{N}\;\left[(B_{ran}\sin\alpha)^2 + 
                 B_{reg,\perp}^2\right] \label{e:ic}\\
  \phi_{Fr} &=& RM \lambda^2 = 0.81 \; n_e\; (B_{ran} 
                \cos\alpha + B_{reg,\pl})\; d \lambda^2
                 \nonumber\\
   P_c      &=& 0.7 \; I_c \nonumber
\end{eqnarray}
where $B_{ran}$ is the constant strength of the random magnetic field
in $\mu$G, $\alpha$ its random angle with the line of sight, $C$ a
proportionality constant, $N$ the number of cells along the line of
sight and $d$ the path length. The total emission from the layer
($\approx CN$) is comparable for different cell sizes, therefore a
factor $1/N$ is added to Eq.~(\ref{e:ic}). The polarized emission in
each cell $P_c$ equals 
the maximum polarization of synchrotron radiation $I_c$ generated in a
cell. For an electron energy power law distribution $N(E) \propto
E^{-\gamma}$, the degree of polarization $p$ is related to the
spectral index $\gamma$ of the electron energy distribution as $p(\gamma) = 
(3\gamma+3)/(3\gamma + 7)$ (Burn 1966). For $\gamma$ around 2.7, the
maximum polarization is $\sim$~70\% of the total intensity. The
polarization angle of the emission generated in each cell
$\phi_{in}$ is taken to be perpendicular to the position angle of the
perpendicular magnetic field. The position angle of the random
magnetic field component $\phi_r$ is random, and that for the regular
component is chosen in the direction of Galactic longitude. Therefore
the polarized intensity emerging from a cell is
\begin{equation}
  {\mathbf P_c} = 0.7 I_c \; {\rm e}^{-2i(\phi_{Fr} + \phi_{in})} + 
                  0.7 I_b \; {\rm e}^{-2i(\phi_{Fr} + \phi_b)}
\end{equation}
for a cell that is irradiated with polarized intensity $I_b$ and
polarization angle $\phi_b$. The input and output parameters are given
in Table~\ref{t3:mpar}.

\section{Estimate of the parameter $C$}
\label{a:c}

The total power per unit volume per unit frequency of
synchrotron emission is (Rybicki \& Lightman 1979)  
\begin{eqnarray}
  P_{tot}(\omega) &=& \frac{\sqrt{3}q^3 \kappa^{\prime} B\sin\alpha}
                   {2\pi mc^2(p+1)}\,
  \Gamma\left(\frac{p}{4}+\frac{19}{12}\right)
  \Gamma\left(\frac{p}{4}-\frac{ 1}{12}\right)\nonumber \\
  & &  \times\left(\frac{mc\omega}{3qB\sin\alpha}\right)^{-(p-1)/2}
  \nonumber \\
  &=& \frac{3\sqrt{3}}{16\pi^2}\frac{q^4\kappa^{\prime}B_{\perp}^2}{m^2c^3\nu}
  \Gamma\left(\frac{28}{12}\right)\Gamma\left(\frac{2}{3}\right)
   \,\,\,\mbox{for $p=3$}\label{e:p}
\end{eqnarray}
where $q$ is the electron charge, $\kappa^{\prime}$ is the
proportionality constant in a power law particle spectrum $N(\gamma)$
with spectral index~$p$ 
($N(\gamma)d\gamma=\kappa^{\prime}\gamma^{-p}d\gamma$), $B\sin\alpha =
B_{\perp}$
is the magnetic field component perpendicular to line of sight, $m$ is
the electron mass, $c$ the speed of light and $\omega$ the angular
frequency of the radiation. $\Gamma$ denotes the gamma-function.
 
We can estimate $\kappa^{\prime}$ by assuming that the electron particle
spectrum throughout the ISM is equal to the local value in the solar
neighborhood. Longair (1981) gives a value of $\kappa =
2.9~10^{-5}$~particles~m$^{-3}$~GeV$^{-(1-p)}$ for the proportionality
constant of the particle spectrum as a function of energy
$N(E)dE=\kappa E^{-p}dE$ derived from direct measurements of the
particle spectrum in the local ISM, in agreement with the value
  found by Golden et al.\ (1994). This can be converted into
$\kappa^{\prime}$ as $\kappa^{\prime}=\kappa (mc^2)^{(1-p)}$. Using $p
= 3$ and converting to cgs units gives $\kappa^{\prime} =
1.0~10^{-4}$~part~cm$^{-3}$~ergs$^2$. Inserting this into
Eq.~(\ref{e:p}) yields

\begin{equation}
P_{tot}(\nu) = 4.2~10^{-39}\frac{B_{\perp}[\mu\mbox{G}]^2}{\nu[\mbox{MHz}]}
      \mbox{~W~m$^{-3}$~Hz$^{-1}$} \label{e:ptot}
\end{equation}
This is the volume emissivity of synchrotron emission as a function of
frequency and magnetic field. We can check how reasonable this number
is by comparing to the total power observed from the Caswell (1976)
radio survey at 10~MHz. The average brightness temperature computed by
Caswell corresponds to a volume emissivity of
$\sim3~10^{-39}$~W~m$^{-3}$~Hz$^{-1}$ (Longair 1981), which agrees well
with our value of $P_{tot}(10~\mbox{MHz}) = 0.42~10^{-39}~B_{\perp}[\mu
  \mbox{G}]^2$~W~m$^{-3}$~Hz$^{-1}$ for $B_{\perp} = 3~\mu$G.

The next step is to describe $P_{tot}$ in terms of the observables.
The observables are in Kelvin, whereas $P_{tot}$ is in
W~m$^{-3}$~Hz$^{-1}$. The power that is detected on 1~m$^2$ of antenna
surface is
\[
P_{cell,ant}~[\mbox{W~Hz$^{-1}$~m$^{-2}$}] = \frac{d^3}{4\pi  D^2}
  P_{tot}~[\mbox{W~m$^{-3}$~Hz$^{-1}$}]\\
\]
with $D$ the distance to the cell. This gives the emissivity of a
source in the direction of the observer on 1~m$^2$ of dish in
Jansky's. Then, to convert into Jansky/beam, only consider the part of
the source which fits into a beam. With a spatial resolution of $D
\tan(5^{\prime})$, the area of a beam is $\pi (1/2~D\tan(5^{\prime})^2)$,
and the number of beams that fits into the source is $4d^2/(\pi
D^2\tan^2(5^{\prime}))$. Then the emissivity in one cell per beam is
\begin{eqnarray}
I_c[\mbox{W~Hz$^{-1}$~m$^{-2}$~beam$^{-1}$}] \;=\; \hspace{4cm}\nonumber \\
\frac{d^3}{4\pi D^2}
  \left(\frac{\pi D^2\tan^2(5^{\prime})}{4d^2}\right)
  P_{tot}~[\mbox{W~m$^{-3}$~Hz$^{-1}$}]\nonumber
\end{eqnarray}
so that, using Eq.~(\ref{e:ptot}), the emissivity $I_c$ in Jy/beam is
$I_c = 0.21\; d$[pc]~K using that 1~mJy/beam~$\approx$~0.13~K at
350~MHz. 
Combining this result with Eq.~(\ref{e:ic}) yields an estimate for
$C$: 
\[
C = \frac{0.21 L}{\nu} \approx 0.5
\]
for a path length L~=~900~pc and $\nu = 350$~MHz, where we used $N =
L/d$. We use a value of $C=1$ in the model.

\end{document}